\documentclass[10pt,journal,compsoc]{IEEEtran}
\ifCLASSOPTIONcompsoc
  \usepackage[nocompress]{cite}
\else
  \usepackage{cite}
\fi
\ifCLASSINFOpdf
\else
\fi
\usepackage[leqno]{amsmath}
\usepackage{amssymb}
\usepackage[normalem]{ulem}
\usepackage[]{graphicx}
\usepackage[margin=3cm]{geometry}
\usepackage[]{color}

\usepackage{caption}
\usepackage{subcaption}

\newtheorem{thm}{Theorem}[section]
\newtheorem{lem}[thm]{Lemma}

\newtheorem{cor}[thm]{Corollary}

\newcommand{\N}{N}

\newcommand{\C}{\mathcal C}

\newcommand{\rev}[1]{{\textcolor{black}{#1}}}

\usepackage{xcolor}  

\begin{document}
%
\title{Planar Rooted Phylogenetic Networks}
%
%
%
%

\author{Vincent~Moulton,
        Taoyang~Wu 
\IEEEcompsocitemizethanks{\IEEEcompsocthanksitem V. Moulton and T. Wu are with  School of Computing Sciences, University of East Anglia (UEA), Norwich, NR4 7TJ, UK\protect\\
E-mail:  v.moulton@uea.ac.uk and taoyang.wu@uea.ac.uk
}
}

%
%

\markboth{Transactions on Computational Biology and Bioinformatics}%
{Moulton and Wu: planar networks}
%



\IEEEtitleabstractindextext{%
\begin{abstract}
A \rev{rooted} phylogenetic network is a directed acyclic graph with a single root, 
whose sinks correspond to a set of species.
As such networks are useful for representing the evolution of species that have undergone reticulate evolution, there has been great interest in 
developing the theory behind and algorithms for constructing 
them. 
However,  unlike evolutionary  trees, these networks can be highly non-planar, which can  make them difficult
to visualise and interpret. Here we investigate properties of
planar \rev{rooted} phylogenetic networks and algorithms for deciding whether or not \rev{rooted} networks have certain special planarity properties. 
\rev{In particular, we introduce three natural subclasses of planar rooted  phylogenetic networks and show that they form a hierarchy. 
	In addition, for the well-known level-$k$ networks, we show that level-1, -2, -3 networks are always outer, terminal, and upward planar, respectively, and that level-4 networks are not necessarily planar. Finally, we show that a regular network is terminal planar if and only if it is pyramidal. 
}
Our results make use of the highly developed field of planar digraphs, and
we believe that the link between phylogenetic networks and \rev{planar graphs} should prove useful in future for developing new 
approaches to both construct and visualise phylogenetic networks.
\end{abstract}

}

\maketitle

\IEEEdisplaynontitleabstractindextext

%
\IEEEpeerreviewmaketitle

\section{Introduction}

Phylogenetic networks are a generalization of evolutionary trees
that are commonly used to study species whose evolution have
involved reticulate processes such as recombination, hybridization and lateral gene transfer. 
These networks have been used to analyze the evolution of 
organisms such as viruses, plants,
bacteria, as well as animals (see e.g. \cite{huson}). In general, 
phylogenetic networks come
in two main types: Those that aim to represent evolutionary
patterns in the data, and
those that try to represent the actual evolutionary history (sometimes called
{\em implicit} and {\em explicit} networks, respectively; see
e.g. \cite{morrison}). 
Note that for the special case of evolutionary trees, implicit and explicit 
networks can be thought of as corresponding to unrooted and rooted trees, respectively.
More generally, implicit networks tend to be undirected
graphs whereas explicit ones are directed.

\begin{figure*}[ht]%
	\centering
	\begin{subfigure}{0.41\textwidth}
		\centering
		\includegraphics[width=0.8\linewidth]{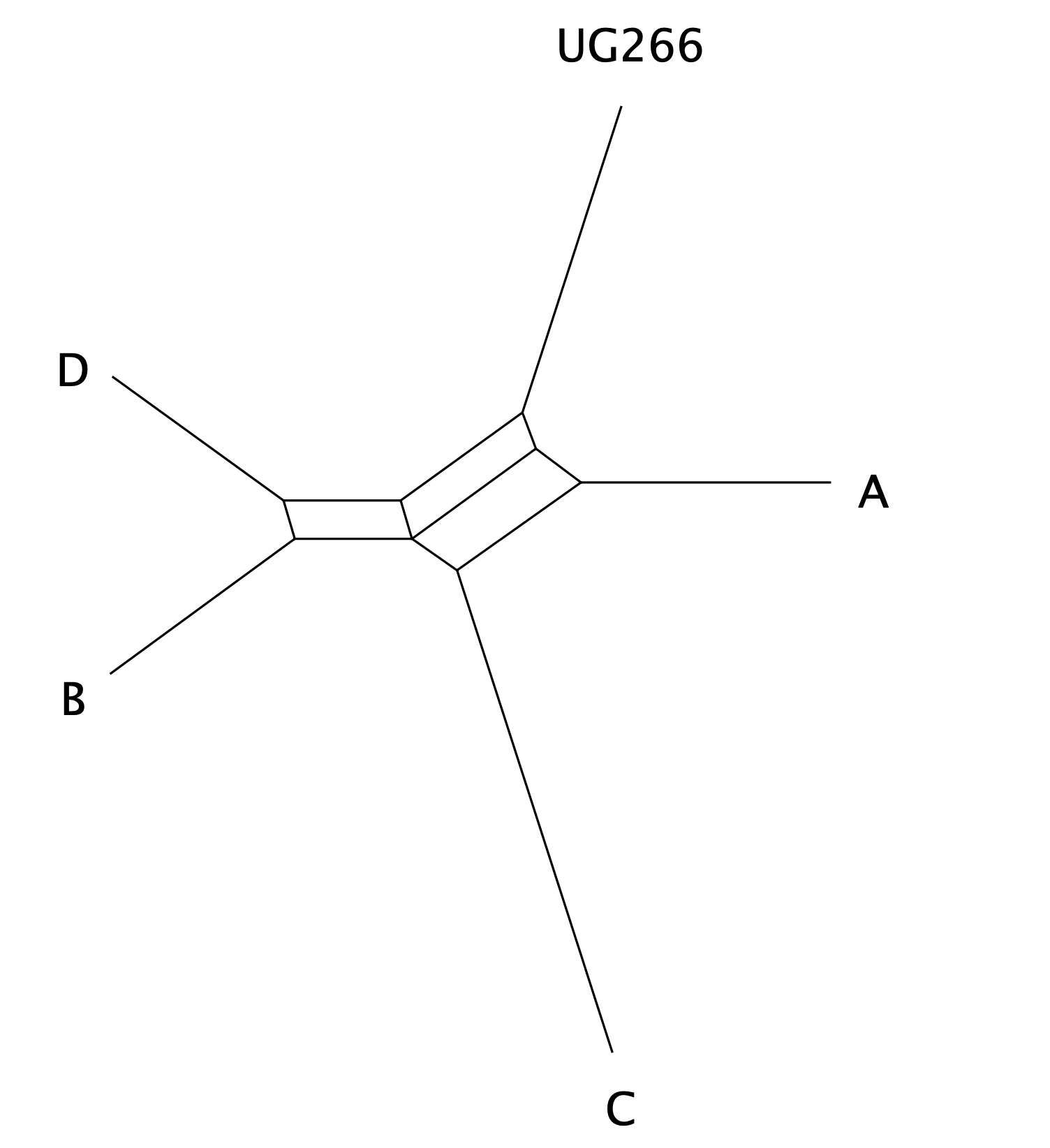}%
	\end{subfigure}
	\begin{subfigure}{0.41\textwidth}
		\centering
		\includegraphics[width=0.4\linewidth]{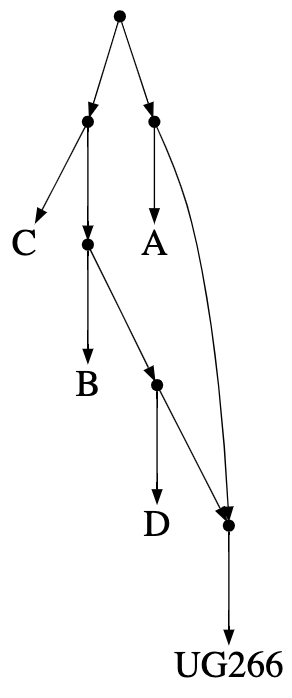}%
	\end{subfigure}
	\caption{Examples of implicit and explicit phylogenetic networks on a set of HIV viruses.
		The labels A--D correspond to viral  
		subtypes, whereas UG266 corrsponds to
		a virus that is an A/D-recombinant.
		Left: A split network generated by Neighbornet; Right: a level-1 network generated by Trilonet.
		Note that directions in the right network make it somewhat clearer that UG266 has arisen as 
		an A/D-recombinant.}%
	\label{fig:net:example}%
\end{figure*}

Although the concept of a phylogenetic network has
been considered in the literature for quite some
time, the first algorithmic approaches to produce such networks
mainly focused on generating implicit networks.
In particular, one of the archetypal types of implicit networks are the so-called
{\em split networks} that were first introduced in \cite{bandelt}. 
For example, in the left panel of Figure~\ref{fig:net:example} we
present an example of a split network generated for   
collection of HIV virus sequences described in \cite[Chapter 21]{lemey}
that was generated by the Splitstree program \cite{husbry} using the NeighborNet algorithm \cite{brymou}.
In general, split networks are isometric subgraphs of hypercubes \cite{bandelt}, and 
which means that they can be potentially highly non-planar (i.e. it is not possible to draw them 
in the plane without some edges crossing one another).
Even so, split networks have proven to be very popular, one
of the main reasons probably being that algorithms such as
the NeighborNet algorithm are designed to generate split networks
that {\em are planar} which greatly facilitates their visualization (see e.g. \cite{drehus,gamhus,spillner}) 
and subsequent biological interpretation. 

More recently, the focus of the phylogenetic networks literature
has shifted more towards explicit networks, \rev{and especially rooted networks.}
Essentially, a \rev{rooted} phylogenetic network (which from now
on we will just call a phylogenetic network) is a directed acyclic
graph with a single source \rev{called the} root,  whose sinks (which 
are usually \rev{called} the leaves of the network) correspond to the set of species. 
For example, in the right panel of Figure~\ref{fig:net:example} we present a
network generated using the Trilonet program \cite{oldman}. 
There are several types of phylogenetic networks (see e.g. \cite[Chapter 10]{S} for a recent review), and
various programs are available for their 
generation (e.g. \cite{hussco,solis,wen} and \cite{wallin} for a recent comparison
of some of these programs).
Even so, although some of these types of networks (such as so-called level-1 networks) are
always planar,  just as with split networks this is not true in general. Thus, it is interesting
to study planarity of phylogenetic networks, with the view to
developing new approaches and algorithms for generating 
phylogenetic networks that can be more readily visualized.

In this paper we shall present some new concepts and 
results concerning planar phylogenetic networks, which we 
now briefly summarize. To help prove our results we
shall use the highly developed theory of planar directed 
graphs and posets, some of which we shall review in the next section. In Section~\ref{sec:planar}, we 
then consider planar phylogenetic networks and introduce and compare 
three subclasses of such networks: upward planar,
terminal planar, 
and outer planar networks. 
Upward and outer planar phylogenetic networks are special examples planar \rev{digraphs} with the same name, 
whereas terminal planar networks are networks where the source and 
sinks (i.e., the root and leaves) must all lie in the so-called outer 
face of some planar drawing of the network. 
We present a characterisation of terminal planar 
networks in Theorem~\ref{thm:upward}, and use this result to show that the
four types of planar networks form a hierarchy (Theorem~\ref{thm:inclusions}). 
As a consequence we also show that there are polynomial time algorithms
for deciding whether or not a phylogenetic network
is contained in each one of these classes or not (Theorem~\ref{thm:algorithm}). 

In Sections~\ref{sec:level} and \ref{sec:regular} 
we investigate planarity for two special classes of phlogenetic
networks: level-$k$ ($k \ge 0$) \cite{J06} and regular networks \cite{BS05}. 
Loosely speaking, the level of a network is measure of how far the network
is from being a tree, where level-0 networks are trees. For level-$k$ networks
we show that level-1, -2, -3 networks are always outer, terminal, and upward planar,
respectively, and that level-4 networks are not necessarily planar
(Theorem~\ref{thm:level}). We also characterise
when level-2 networks are terminal planar (Theorem~\ref{thm:level:two}).  
A regular network can be thought of as being a
Hasse diagram of some collection of clusters of 
a set of species.
We show that a regular network is terminal planar 
if and only if the underlying set of clusters forms a {\em pyramid} \cite{Diday} 
(Theorem~\ref{thm:regular}).
We conclude in Section~\ref{sec:discussion} with a discussion and 
some open problems.

\section{Preliminaries}
\label{sec:pre}

In this paper we shall assume that $X$ is a finite set with $|X|\ge 2$. The
set $X$ usually corresponds to a collection of extant species.

\subsection{Posets and cluster systems}

A set $Y$ with a partial order $\preceq$ (i.e., a binary reflexive, antisymmetric, and transitive relation)
is called a partially ordered set or {\em poset}, 
and denoted by $(Y,\preceq)$. A poset $(Y,\preceq)$ is a {\em lattice} if for each 
pair of elements $(y_1,y_2)$ in $Y$, the set $Y$ contains both 
a join $y_1\vee y_2$ (i.e., a least upper bound) and 
a meet $y_1 \wedge y_2$ (i.e., a greatest lower bound) of $y_1$ and $y_2$. 
Note that in case either a join or a meet exists then  it is necessarily unique. 
The partial order $\preceq$ is referred to as a {\em linear order} if in addition, 
for each pair of distinct elements $y_1$ and $y_2$ in $Y$, we have either $y_1\preceq y_2$ or $y_2\preceq y_1$ 
(but not both). 
Clearly, a linear order is 
necessarily a  lattice.  Given a poset $P$, we let $P^*$ be $P$ if it has the 
bottom element (i.e. a least element), and $P$ with an extra bottom element 
$\hat{0}$ added otherwise.

The {\em Hasse diagram} $H(P)$ of a poset $P=(Y,\preceq)$ is the digraph 
with vertex set $Y$ and so that $(y_1,y_2)$ is a directed edge from 
$y_1$ to $y_2$ in $H(P)$ if $y_1$ {\em covers} $y_2$, that is, $y_2\preceq y_1$ and 
there is no element $y_3\in Y-\{y_1,y_2\}$ with $y_2 \preceq y_3 \preceq y_1$. 
Note that the direction of the edges in our definition of the Hasse diagram is 
chosen for the convenience later on, and is the opposite 
to the way commonly used in the poset literature. 
However, this difference will be immaterial for the results presented here.  
It is well-known in poset theory that each partial order $\preceq$ on $Y$ 
has a {\em linear extension}, that is, a linear order $\preceq_1$ on $Y$ 
such that $y_1\preceq y_2$ in $Y$ implies that $y_1\preceq_1 y_2$.
Following~\cite{D41}, the (Dushnik and Miller) {\em dimension} of a poset $(Y,\preceq)$ is 
the smallest cardinal number $m$ such that $\preceq$ is the intersection 
of $m$ linear extensions $\preceq_1,\dots,\preceq_m$ of $\preceq$, 
that is, $y_1 \preceq y_2$ if and only if $y_1 \preceq_i y_2$ for each $1\le i \le m$.

A {\em cluster system $\C$ (on $X$)} is a collection of non-empty subsets of $X$ that contains $X$ and 
the subset $\{x\}$ for each $x\in X$, where each 
subset of $X$ in the system is referred to as a {\em cluster}. 
Furthermore, $\C$ is {\em closed under intersection} if 
for each pair of sets $A$ and $B$ in $\C$, we have either $A\cap B=\emptyset$ or $A\cap B\in \C$.
Note that any cluster system $\C$ is a poset $P(\C)=(\C,\subseteq)$ 
under the ordering induced by taking set-inclusion; we  
let $P^*(\C)$ be the poset obtained from $P(\C)$ by adding 
the emptyset $\emptyset$ as the bottom element. 
We now make a simple observation that we will use later on. We 
include a proof for completeness.

\begin{lem}
	\label{lem:closed:lattice}
	Suppose that $\C$ is a cluster system on $X$ that is 
	closed under intersection. Then the poset $P^*(\C)$ is a lattice. 
\end{lem}

\begin{IEEEproof}
	Let $A$ and $B$ be two distinct 
	elements in $\C$. Let $A\wedge B$ be $\emptyset$ if $A$ and $B$ 
	are disjoint, and $A\cap B$ otherwise. Since $\C$ is closed under intersection, it 
	follows that $A\wedge B$ is contained in $P^*(\C)$ and 
	it is in fact the meet of $A$ and $B$.
	
	Next, let $u(A,B)$ be the collection of subsets $C$ in $\C$ 
	such that $A\subseteq C$ and $B\subseteq C$. Then $u(A,B)$ is 
	non-empty as it contains $X$. Now, for $C_1$ and $C_2$ in $u(A,B)$, 
	we have  $A\cup B\subseteq C_1\cap C_2$, and 
	hence $C_1\cap C_2\not =\emptyset$. Therefore, $C_1\cap C_2 \in u(A,B)$. In other words, 
	$u(A,B)$ has a minimum element $C_0$ such 
	that $C_0\subseteq C$ for all $C\in u(A,B)$. 
	Clearly, $C_0$ is the join of $A$ and $B$ in $P(\C)$, and hence 
	also their join in $P^*(\C)$. Hence $P^*(\C)$ is a lattice. 
\end{IEEEproof}	

\subsection{Planar graphs}

We shall assume that all sets and graphs are finite,
and that all graphs are simple, that is, 
they contain neither loops nor parallel edges.
For integers  $m,n\ge 1$, let $K_m$ be the complete graph
with $m$ vertices, and $K_{m,n}$ be the complete bipartite graph with $m$ 
vertices in one set and $n$ vertices in the other set.

We now recall some standard definitions for planar graphs and graph drawing, 
largely following~\cite{Diestel} and~\cite{Tamassia13}.
A {\em drawing} $\Gamma=\Gamma(G)$ of a graph $G$ maps each vertex $v$ to a 
distinct point $\Gamma(v)$ of the Euclidean plane (endowed with the $x$ and $y$ coordinates) 
and each edge $(u,v)$ to an {\em arc} $\Gamma(u,v)$, that is, a simple open 
Jordan curve with endpoints $\Gamma(u)$ and $\Gamma(v)$. 
For clarity, the points $\Gamma(v)$ corresponding to a vertex $v$ in $G$ 
will be referred to as a node in the drawing, to differentiate  
them from the points inside of the arcs. 
Such a drawing is called a {\em straight-line drawing} if each arc in the drawing 
is a straight line segment between its endpoints.
A drawing is {\em planar} is no two distinct arcs intersect except, 
possibly, at common endpoints. A graph is {\em planar} if it admits a planar drawing.
Note that such a drawing partitions the plane into connected regions called {\em faces}. 
There is precisely one unbounded face of $\Gamma$, referred to as the {\em outer-face} of $\Gamma$.
If a planar graph has a drawing in which every vertex is contained in the outer-face, then
the graph is called {\em outer planar}.

Given two graphs $G$ and $G'$, we say
$G'$ is a {\em minor} of $G$ if $G'$ can be obtained from $G$ 
by repeated deletion (of vertices and edges) and contractions. Furthermore, 
we say $H$ is a subdivision of $G$ if  $H$ is obtained from $G$ by replacing the 
(directed) edge of $G$ with (directed) paths between their ends (so that none of
these paths has an inner vertex on another path or in $G$). 
The well-known Kuratowski theorem states that a graph $G$ is planar if and 
only if none of the subgraph of $G$ is a subdivision of either $K_5$ or $K_{3,3}$;  
equivalently, the Wagner \rev{theorem} states that 
$G$ is planar if and only if $G$ contains neither $K_5$ nor $K_{3,3}$ 
\rev{as} a minor (see, e.g.~\cite[Theorem 4.4.6]{Diestel} for a proof of both theorems). 
Moreover, $G$ is outer planar if and only if no subgraph of $G$ is a 
subdivision of either $K_{4}$ or $K_{2,3}$; see, e.g.~\cite[p.117]{Br99}.

Note that if $G$ is a digraph, then each edge $(u,v)$ will be mapped to a directed 
arc from $\Gamma(u)$ to $\Gamma(v)$. Clearly, a digraph $G$ is planar if 
and only if its underlying graph $U(G)$ (e.g. the one obtained from $G$ by forgetting the directions of edges) 
is planar. Furthermore, a planar drawing of a digraph is called {\em upward} if each arc  
is a curve of monotonically nondecreasing in the $y$-coordinate.  As noted in~\cite{G95}, 
a digraph $G$ is {\em upward planar} 
if it admits a straight-line upward planar drawing. Another useful result concerning 
upward-planar graphs is as follows.
An {\em st-digraph} is an acyclic digraph with a single source, denoted by $s$, and a single sink, 
denoted by  $t$, and the arc $(s,t)$. Then $G$ is upward-planar if and only 
if it is a spanning subgraph of a planar $st$-digraph, that is, directed edges 
can be added to $G$ such that the resulting digraph is a planar $st$-digraph~\cite[Theorem 1]{G95}.

\section{Planar phylogenetic networks}
\label{sec:planar}

In the following we shall mainly use the definitions for
phylogenetic networks presented in~\cite[Section 10]{S}.
A {\em  phylogenetic network on $X$}, or simply a phylogenetic 
network when the underlying set $X$ is clear from the context,  is 
an acyclic directed graph $N=(V,A)$ with $X \subseteq V$, 
a single source $\rho_N$ (the root), $X$ the set of sinks in $N$, 
and no vertices with indegree 1 and outdegree 1. 
In particular, each edge is directed away from the root of $N$.
Two phylogenetic networks $\N=(V,A)$ and $\N'=(V',A')$ on $X$ are {\em isomorphic} 
if there exists a bijective map $f: V \to V'$ such that $(u,v)\in A$ if and only if $(f(u),f(v))\in A'$, 
and for all $x\in X$, $f(x)=x$.

\begin{figure}[ht]
	\center
	\scalebox{0.5}{\includegraphics{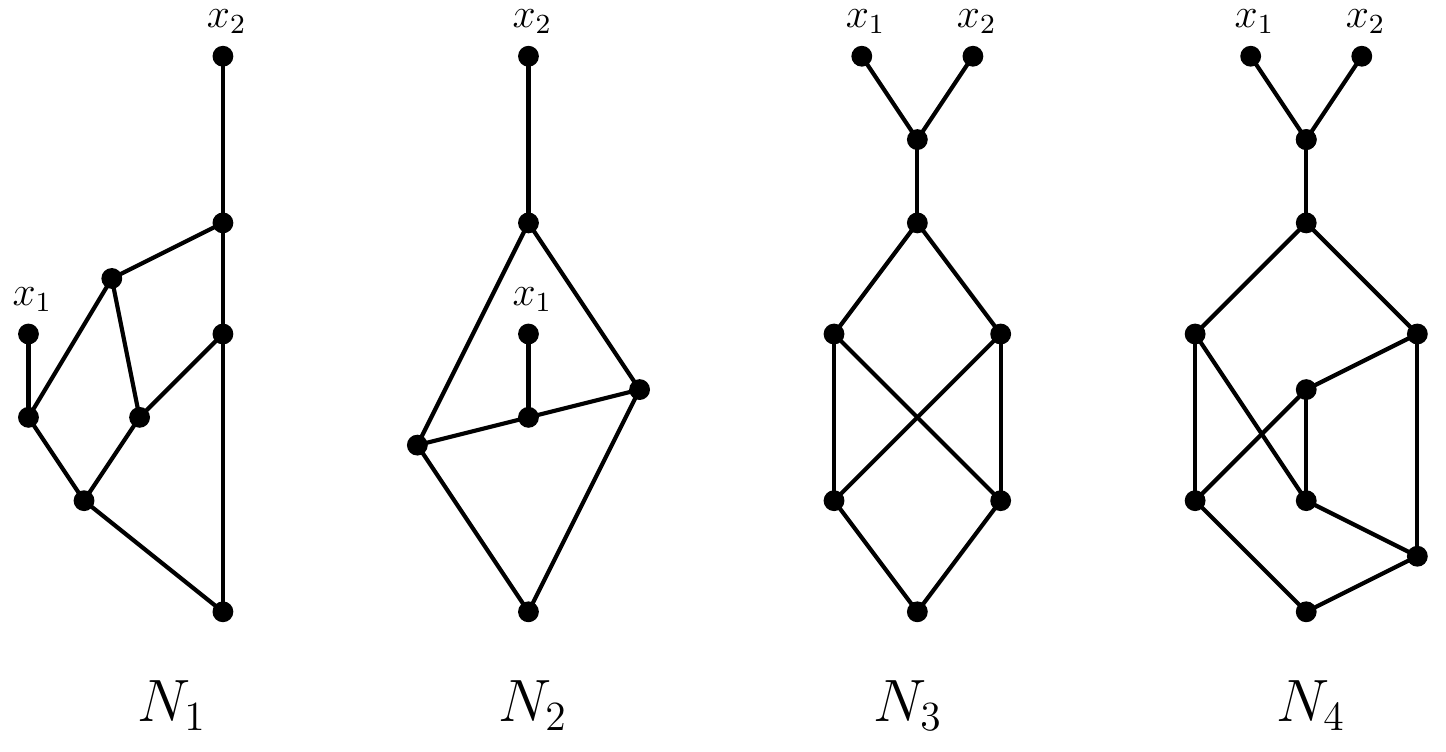}}
	\caption{Examples of phylogenetic networks on $X=\{x_1,x_2\}$.   
		$N_1$: terminal planar but not outer planar,
		$N_2$: upward planar but not terminal planar,
		$N_3$: planar but not upward planar,
		$N_4$: not planar. 	
		The direction of each edge is pointed upward, and omitted here for 
		simplicity; a convention used for all directed graphs in this paper \rev{from here on}. }
	\label{fig:examples}
\end{figure}

A phylogenetic network on $X$ is called {\em planar, outer planar or upward planar} 
if as a digraph it satisfies each of these properties.  
Furthermore, we defined such a network to be
{\em  \rev{terminal} planar\footnote{We were unable to find an agreed name
		for a vertex in an arbitrary digraph that is either 
		a source or  a sink, so for phylogenetic networks
		we call such a vertex a {\em terminal} of the network.}} if admits a planar drawing 
in which the set $\{\rho_N\} \cup X$ is contained in the outer-face
(note that this is analogous to {\em outer-labelled split networks} \cite{drehus}, where 
the elements in $X$ must all lie in the outer face of the network).
Clearly, if a phylogenetic network is outer planar, then it is terminal planar.
Also, note that a phylogenetic network $N$ is outer planar (respectively terminal planar or planar) if and 
only if its underlying graph $U(N)$ is  outer planar (respectively terminal planar or planar). 
In Fig.~\ref{fig:examples} we present some examples of these concepts;
we shall prove that the networks in this figure have the stated properties
in Theorem~\ref{thm:inclusions} and Theorem~\ref{thm:level} below.

We now present a result relating terminal planar and upward planar phylogenetic networks on $X$.
Given a phylogenetic network $N=(V,E)$ on $X$, we let $N^*$ be the digraph 
with vertex set $V \cup \{t\}$ for some $t \not\in V$ which will be referred to as the {\em sink vertex}, 
and directed edge set $A \cup \{(x,t) \,:\, x \in X\}$. We call $N^*$ 
the {\em completion} of $N$. 

\begin{thm}
	\label{thm:upward}
	Suppose that $N$ is a phylogenetic network on $X$. Then $N$ is terminal planar if and only if  
	the completion $N^*$ of $\N$ is upward planar.
\end{thm}

\begin{IEEEproof}
	Suppose first that $N^*$ is upward planar. Fix an upward drawing $\Gamma^*$ of $N^*$. 
	Consider the drawing $\Gamma$ of $N$ obtained from $\Gamma(\N^*)$ by 
	removing the sink $\Gamma^*(t)$ and the arcs incident with it. Then $\Gamma$ is a planar drawing of 
	$N$ such that the set $\{\rho_N\} \cup X$ is contained in the outer-face of $\Gamma$.	
	So $N$ is terminal planar.
	
	Conversely, suppose that $N$ is terminal planar. Fix a planar drawing $\Gamma$ of
	$N$ in the plane such that the set $\{\rho_N\} \cup X$ is contained in the outer-face. 
	Without loss of generality, we may assume that the $y$-coordinate of each point
	in the drawing $\Gamma$ is strictly positive. 
	Take a point $w$ in the interior of the outer-face in the 
	plane whose $y$-coordinate is negative. Join
	this point to each element $x \in X$ with a curve $\gamma_{w,x}$ 
	contained in the outer-face such
	that $\gamma_{w,x} \cap \gamma_{w,x'} = \{w\}$ for all
	$x, x' \in X$. Then add in an arc between the $\rho_N$ and
	$w$ which does not intersect any existing arcs, except possibly in $w$ or $\rho_N$.
	This gives a planar drawing  $\Gamma'$ of the  digraph 
	$$
	N'= (V \cup \{t\}, A \cup \{(x,t) \,:\, x \in X\} \cup \{(\rho_N,t)\})
	$$
	in which $t$ is mapped to $w$ under $\Gamma'$. 
	So $N'$ is a planar $st$-digraph (with source $s=\rho_N$ and sink $t$). 
	It follows that $N^*$ is upward planar by \cite[Theorem 1]{G95} that we stated above.
	Finally, as $N$ is a subgraph of $N^*$, it follows that 
	$N$ is also upward planar. 
\end{IEEEproof}

Let $\mathcal P_{p}(X)$, $\mathcal P_{u}(X)$, $\mathcal P_{t}(X)$, $\mathcal P_{o}(X)$ denote
the classes of planar, upward planar, terminal planar, and 
outer planar phylogenetic networks on $X$, respectively. 
We now show how these classes are related to one another.

\begin{thm}
	\label{thm:inclusions}
	For any $X$, $|X|\ge2$, we have 
	$$
	\mathcal P_{o}(X) \subsetneq \mathcal P_{t}(X) \subsetneq \mathcal P_{u}(X) \subsetneq \mathcal P_{p}(X). 
	$$
\end{thm}

\begin{IEEEproof}
	By definition we have $\mathcal P_{o}(X) \subseteq \mathcal P_{t}(X)$ 
	and $\mathcal P_{u}(X) \subseteq \mathcal P_{p}(X)$. 
	Furthermore, $\mathcal P_{t}(X) \subseteq \mathcal P_{u}(X)$ follows 
	from Theorem~\ref{thm:upward}. 
	
	We now show that all of the set inclusions in the statement of the theorem  
	are strict, for the cases $Y=\{x_1,x_2\}$ and $X=\{x_1,x_2,\dots,x_t\}$, $n\ge 3$,  
	which will complete the proof of the theorem.
	
	\begin{figure}
		\center
		\scalebox{0.5}{\includegraphics{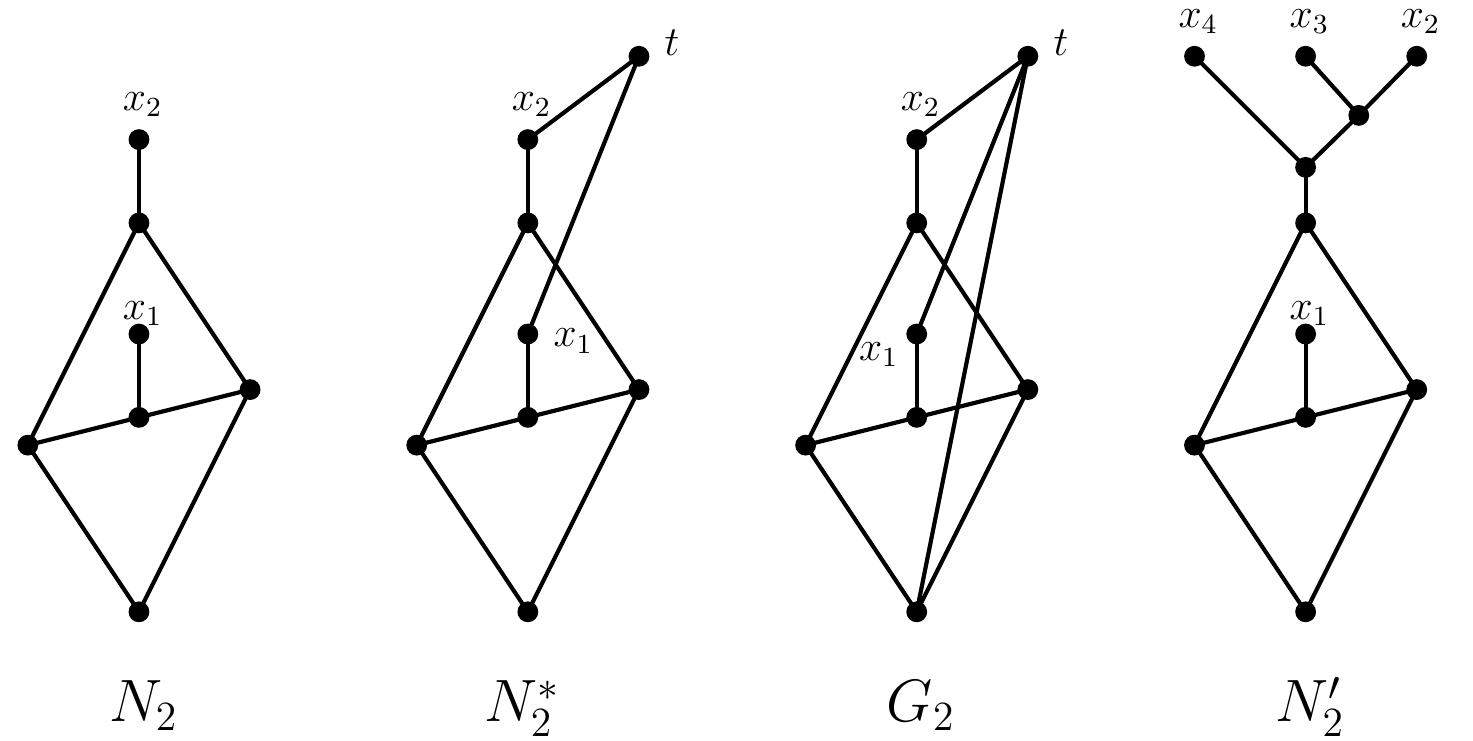}}
		\caption{  Four graphs in the proof of Theorem~\ref{thm:inclusions}. }
		\label{fig:case1} 
	\end{figure}
	
	First, consider the network $N_2$ on $Y$ in Fig.~\ref{fig:examples}
	which is clearly contained in $\mathcal P_{u}(Y)$.  We
	claim that $N_2 \not\in \mathcal P_{t}(Y)$. Indeed, if this were not the case, 
	then $N_2$ is terminal planar, and hence its completion $N_2^*$ is upward planar 
	in view of Theorem~\ref{thm:upward}. Therefore, $N_2^*$ is  a spanning 
	subgraph of an $st$-digraph denoted by $\widetilde{N}_2^*$. 
	This implies that the graph $G_2$ in Fig.~\ref{fig:case1} 
	is a minor of $U(\widetilde{N}_2^*)$. However, $G_2$ contains $K_{3,3}$ as 
	one of its minors, and hence \rev{is} not planar, a contradiction. 
	Now, let $T_n =T(x_2,\dots,x_n)$ be the 
	binary tree on $\{x_2,\dots,x_n\}$ given by taking the Hasse diagram of
	the cluster system that contains all clusters of the form $\{x_2,\dots,x_k\}$, $3\le k \le n$, 
	and  construct the  phylogenetic network $N'_2$ on $X$ by replacing
	$x_2$ with the tree $T(x_2,\dots,x_n)$ (see  Fig.~\ref{fig:case1}  
	for an example with $n=4$. 
	Since both $N_2$ and $T(x_2,\dots,x_n)$ are upward planar, 
	it is straight-forward to construct an upward planar drawing for $N'_2$. 
	A similar to the proof that $N_2$ is not terminal planar
	then implies that $N'_2 \in \mathcal P_{u}(X)\setminus \mathcal P_{t}(X)$.  
	
	Next, consider the network $N_1$ on $Y$ in Fig.~\ref{fig:examples}
	which is clearly contained in $\mathcal P_{t}(Y)$. Furthermore, 
	$N_1$ is not outer planar because $U(N_1)$ contains a 
	subdivision of $K_{2,3}$. Now we  
	construct the phylogenetic network $N'_1$ on $X$ by replacing $x_2$ with 
	the tree $T_n$ as in the last paragraph.  Since $T_n$ is 
	outer planar and $N_1$ is terminal planar, we start with a planar drawing of $N_1$ whose outer-face contains the root and leaves of $N_1$ and then naturally 
	extend it to a planar drawing of $N'_1$ whose outer-face contains the root and leaves of $N'_1$.  
	It follows that $N'_1$ is terminal planar. 
	Furthermore, $N'_1$ is not outer planar because $U(N'_1)$ 
	again contains a subdivision of  $K_{2,3}$. Hence 
	$N'_1 \in \mathcal P_{t}(X)\setminus \mathcal P_{o}(X)$.
	
	\begin{figure}[ht]
		\center
		\scalebox{0.5}{\includegraphics{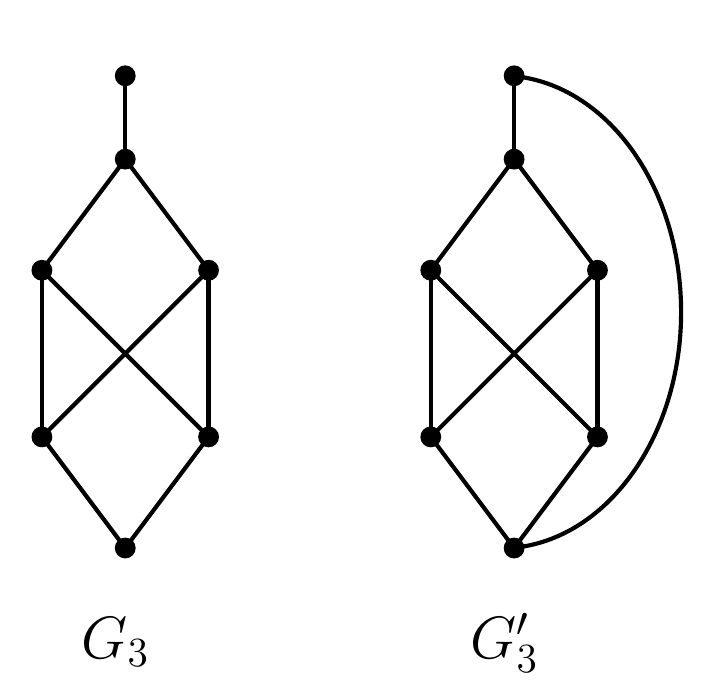}}
		\caption{ Two graphs in the proof of Theorem~\ref{thm:inclusions}.}
		\label{fig:case2}
	\end{figure}
	
	Finally, we claim that 
	$N_3$ in Fig.~\ref{fig:examples}, which is easily seen to be planar, is 
	not in $\mathcal P_{u}(Y)$. If not, then 
	the directed graph $G_2$ (see Fig.~\ref{fig:case2}) obtained from $N_3$ by 
	removing the sinks $x_1,x_2$ and the directed 
	edges incident with them is also upward planar. 
	That is, $G_3$ is a spanning subgraph of an  $st$-digraph $\widetilde{G}_3$ and 
	hence $G'_3$  in  Fig.~\ref{fig:case2}  is a minor of $U(\widetilde{G}_3)$. However, $G'_3$ 
	contains $K_{3,3}$ as its minor, a contradiction. 
	We now  construct a phylogenetic network $N'_3$ on $X$ by replacing
	$x_2$ with the tree $T_n$ as above.   Since $T_n$ is planar 
	and it can be drawn in an arbitrary small neighborhood 	
	of $x_2$,  it follows that 	$N'_3$ is also planar. An
	argument similar to the proof of that $N_3$ is not upward 
	planar now implies $N'_3 \in \mathcal P_{p}(X)\setminus \mathcal P_{u}(X)$.
\end{IEEEproof}


We  now consider some algorithmic aspects of planar phylogenetic
networks. Our main result will follow more-or-less 
directly from some results in the literature
concerning planar digraphs which we now briefly review.
Suppose that $G$ is a digraph with $m$ vertices. 
Then it can be checked in $O(m)$ time if
$G$ is planar/outer planar,
and a straight-line-planar/outer-planar drawing 
for $G$ can be produced in $O(m)$ time in case it is (see
\cite{H74}/ \cite{Mitchell},\cite{deng07}).
Similarly, there are also $O(m)$ algorithms for checking upward planarity and 
producing an upward planar drawing for $G$ in case $G$ has a single source (see \cite{B93}, and 
also \cite[Theorem 15]{G95}), although
the decision problem for deciding 
whether or not a digraph is upward planar is NP-complete in general  \cite{G01}. 

\begin{thm}
	\label{thm:algorithm}
	Suppose $N$ is a phylogenetic network on $X$ with $m$ vertices.
	Then it can be checked in $O(m)$ time whether or not $N$ is 
	planar/upward planar/outer planar/terminal planar, and a
	planar drawing with the corresponding property can be 
	produced in $O(m)$ time if $N$ satisfies any of these properties.
	In particular, if the number of vertices in $N$ is bounded by
	some polynomial $p$ in $|X|$, then we can 
	check if $N$ satisfies any of these planarity properties 
	and if so produce a corresponding planar drawing in $O(p(|X|))$-time.
\end{thm}

\begin{IEEEproof}
	The results concerning planar, upper planar, or outer planar follow from the results 
	mentioned before the statement of the theorem.
	
	To check whether $N$ is terminal planar, we can construct its completion $N^*$ which contains 
	$m+1$ vertices in $O(m)$ time. Since it can be checked in $O(m)$ time
	whether or not $N^*$ is upper planar,  by Theorem~\ref{thm:upward} we can check whether $N$ is 
	terminal planar in $O(m)$ time.  Furthermore, when $N$ is terminal planar, 
	a terminal planar drawing of $N$ can be obtained from an upper planar 
	drawing of $N^*$ in $O(m)$ time using the procedure in  the proof of Theorem~\ref{thm:upward}.
\end{IEEEproof}	


Using Theorem~\ref{thm:algorithm}, we can now 
say something more about checking planarity for 
some well-known classes of phylogenetic networks.
We begin by recalling some definitions (see e.g. \cite[Chapter 10]{S}).
A vertex in a phylogenetic network is either a {\em tree vertex} 
(i.e., a vertex of indegree 1 or the root vertex) or 
a {\em reticulation vertex} (i.e., a vertex of indegree 2 or more).
Furthermore, a phylogenetic network is called {\em binary} if the 
root has outdegree 2, the leaves have indegree 1, and 
all of the all other vertices have indegree 1 and outdegree 2 or 
indegree 2 and outdegree 1.

Now, a {\em tree-child network} is a phylogenetic network for 
which every non-leaf vertex has a child that is a tree vertex (note
a vertex $v$ in a phylogenetic network $N$ is a child of a vertex $u$ in $N$ 
if $(u,v)$ is an arc in $N$). A vertex $v$ is said to be {\em visible} if there is 
a leaf $x_v\in X$ so that every directed path from the root 
to $x_v$ passes through $v$.  A {\em reticulation-visible network} 
is a phylogenetic network for which each reticulation vertex is visible. 
Since a phylogenetic network is tree-child if and only if each vertex is visible, 
it follows that a tree-child network is necessarily reticulation-visible. 
In particular, the corollary below concerning reticulation-visible networks
also holds for tree-child networks.
Finally, a phylogenetic network is {\em normal} if it is tree-child and has 
it does not contain any arc $(u,v)$ such that there is 
another directed path from $u$ to $v$ (these are know 
as {\em redundant} arcs). 

It is known that a binary
reticulate-visible network on $X$ has $O(|X|)$ vertices 
(see e.g. \rev{\cite[Theorem 10.10]{S}}), 
and that a normal network on $X$ has 
$O(|X|^2)$ vertices (see \rev{\cite[Proposition 10.11]{S}}, and the discussion directly after). 
Hence by Theorem~\ref{thm:algorithm} we immediately see that 
the following holds:

\begin{cor}
	Suppose that $N$ is a phylogenetic network on $X$. If $N$ 
	is binary and reticulate-visible, or normal \rev{(not necessarily binary)}, then 	it can be checked in $O(|X|)$ or $O(|X|^2)$ time, respectively,  
	\rev{whether} $N$ is planar, upper planar, outer planar or terminal planar and 
	in case $N$ satisfies any of these properties, a
	planar drawing can be produced with the corresponding property in the same \rev{asymptotic} time.
\end{cor}

\noindent
{\bf Remark:} There are non-binary, terminal planar, tree-child networks on a set $X$ 
that do not have a polynomial number of vertices in $|X|$ (see e.g. \cite[Fig. 10.5 (ii)]{S}).

\section{Level-$k$ networks}
\label{sec:level}

In this section, we shall consider
planarity of level-$k$ networks \cite{J06}, which are defined as follows.
A biconnected component in a graph is a maximal subgraph that does not contain any cut vertex
(i.e. a vertex whose removal disconnects the graph). 
A phylogenetic network is {\em level-$k$}, $k \ge 0$, 
if it is binary 
and it can be converted into a tree by deleting
at most $k$ directed edges from each biconnected component \cite[p.247]{S}. 
Clearly, a level-$0$ phylogenetic network is necessary a tree, 
and a binary phylogenetic network $N$ is level-$k$ if 
and only if $U(N)$ can be converted into a tree by deleting
at most $k$ edges from each biconnected component of $U(N)$.
We let $\mathcal L_{k}(X)$ denote the set of level-$k$ phylogenetic networks on $X$ for $k \ge 0$, so
that $\mathcal L_{i}(X) \subseteq \mathcal L_{k}(X)$ clearly holds for all $0 \le i \le k$.

To prove the first main result of this section
we will use the following observation~\cite{hutton1996upward}: A connected single-source digraph is 
upward planar if and only if all of its biconnected components are upward 
planar (see also~\cite[Lemma 4]{G95}). Since each phylogenetic network $N$ on $X$ is 
a connected and single-source, it follows that $N$ is upward planar if 
and only if all the biconnected components of $N$ are upward planar. 

\begin{thm}
	\label{thm:level}
	The following statements hold:
	\begin{itemize}
		\item[(i)] All level-1 networks are outer-planar, that 
		is, $\mathcal L_1(X)\subseteq \mathcal P_{o}(X)$.
		\item[(ii)] All level-2 networks are upward planar, but some level-2 networks are  
		not terminal planar, that is, $\mathcal L_2(X)\subseteq \mathcal P_{u}(X)$
		and $\mathcal L_2(X)\setminus \mathcal P_{t}(X)\not = \emptyset$. 
		\item[(iii)] All level-3 networks are planar, but some level-3 network are not 
		upward planar, that is $\mathcal L_3(X)\subseteq \mathcal P_{p}(X)$ and
		$\mathcal L_3(X)\setminus \mathcal P_{u}(X)\not = \emptyset$.
		\item[(iv)] Some level-4 networks are non-planar, that is, 
		$\mathcal L_4(X)\setminus \mathcal P_{p}(X)\not = \emptyset$.
	\end{itemize}
\end{thm}

\begin{IEEEproof}
	(i) Assume that $N$ is a level-1 network on $X$ and consider its underlying 
	undirected graph $U(N)$. 
	Since each non-trivial biconnected component of $U(N)$ is a cycle, it follows that none 
	of the subgraphs of $U(N)$ is a subdivision of either $K_{4}$ or $K_{2,3}$, and 
	hence $U(N)$ is outer planar. Thus $N$ is also outer planar. 
	
	\begin{figure}[ht]
		\center
		\scalebox{0.5}{\includegraphics{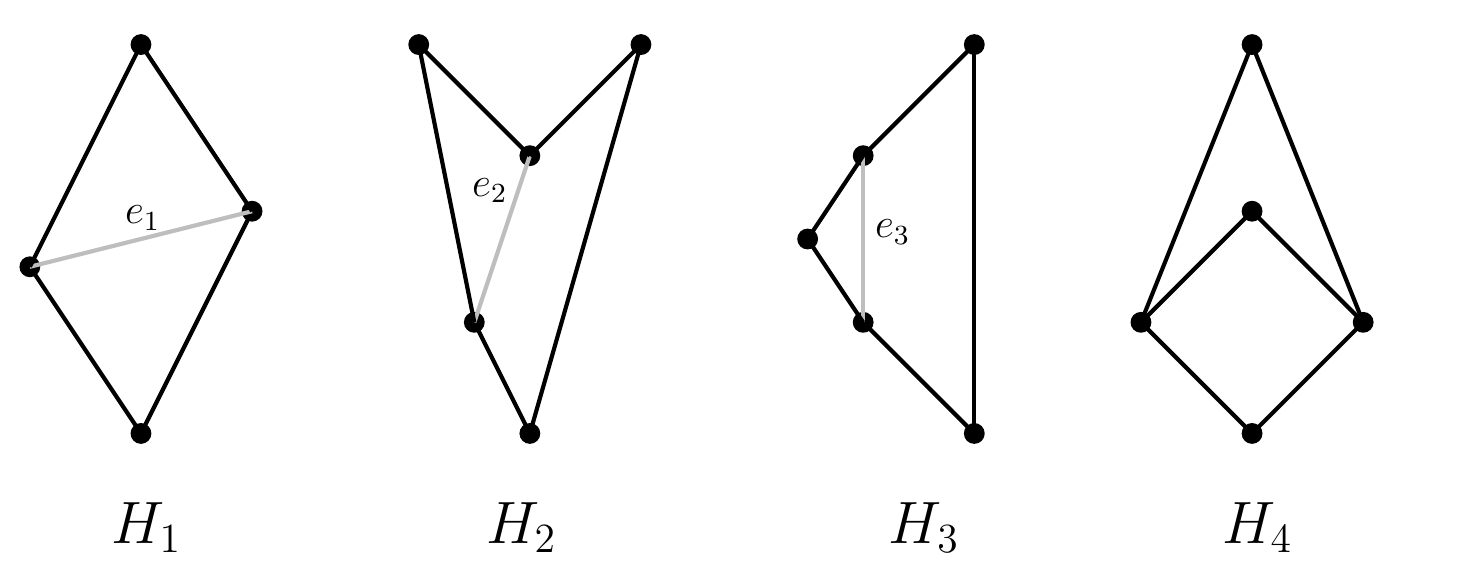}}
		\caption{ The four digraphs for level-2 networks in the proof of~Theorem~\ref{thm:level}; modified from~\cite[Fig. 5]{IM14}. The three edges highlighted in grey and labelled as $e_1$, $e_2$, $e_3$ are referred in the proof of Theorem~\ref{thm:level}(ii). }
		\label{fig:generator}
	\end{figure}
	
	(ii) Assume that $N$ is a level-2 network on $X$. To show that $N$ is upward planar,  it 
	is sufficient to establish the claim that each biconnected component of $N$ is upward planar. 
	To this end, consider  an arbitrary biconnected component $H$ of $N$. Without loss of 
	generality we may further assume that $H$ contains more than one cycle as otherwise 
	the claim clearly holds. Using the results in~\cite[Section 4]{IM14} concerning the 
	generators of level-2 networks, it follows that $H$ is a subdivision of one of the 
	four digraphs in Fig.~\ref{fig:generator}. Since each of these four 
	digraphs in Fig.~\ref{fig:generator} is upward planar, it follows that $H$ is also upward planar, as required. 
	
	Finally, note that the network $N'_2$ constructed in the proof of Theorem~\ref{thm:inclusions}
	is clearly contained in $\mathcal L_2(X)\setminus \mathcal P_{t}(X)$. 
	
	(iii) First, we shall show that $\mathcal L_3(X)\subseteq \mathcal P_{p}(X)$. If not, 
	consider  a level-3 phylogenetic network $N$ on $X$ such that $N$ is not planar. Then $U(N)$
	contains either $K_{3,3}$ or $K_{5}$ as its minor. 
	We shall only consider the case where $U(N)$ contains $K_{3,3}$ as a minor; the proof
	for $K_{5}$ a minor is similar.  In this case, there exists a biconnected component 
	$B$ of $U(N)$ such that $B$ contains $K_{3,3}$ as a minor. 
	This implies that we need to delete at least $4$ edges from $B$ 
	to convert it into a tree, in contradiction to the assumption that $N$ is level-3. 
	
	Finally note that for the network $N_3'$ constructed in the proof of Theorem~\ref{thm:inclusions}
	we clearly have $N'_3\in  \mathcal L_3(X)\setminus \mathcal P_{u}(X)$.
	
	\begin{figure}[ht]
		\center
		\scalebox{0.5}{\includegraphics{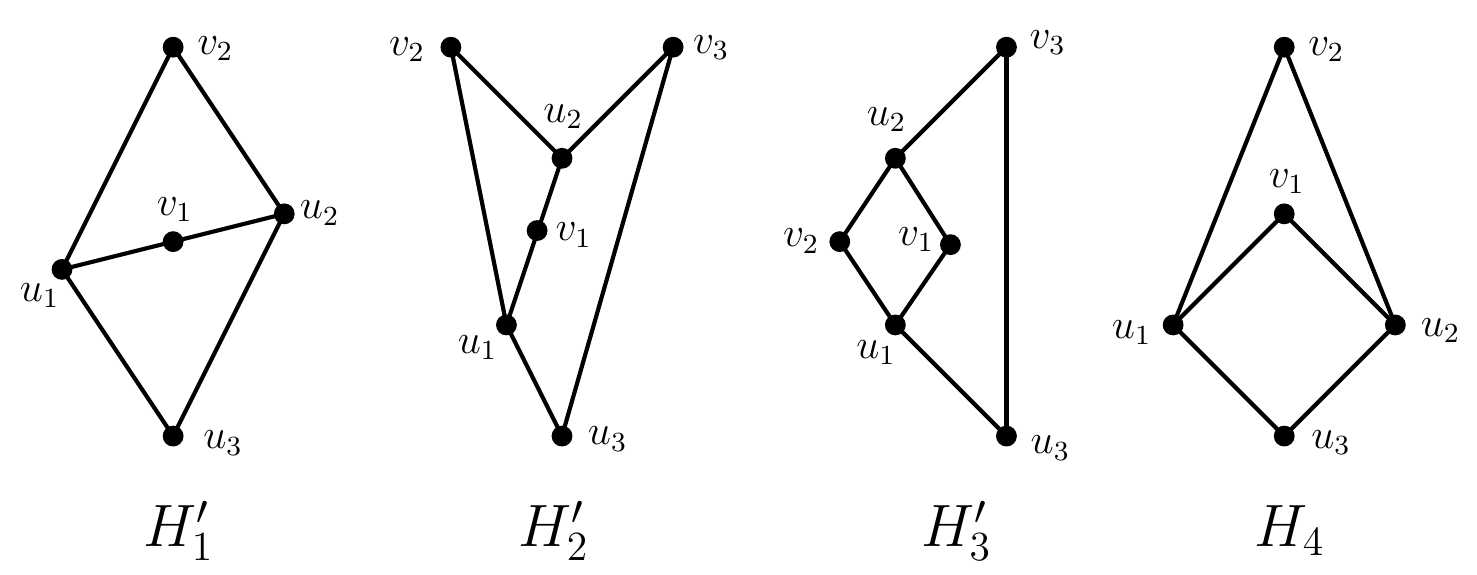}}
		\caption{ The four digraphs 
			in~Theorem~\ref{thm:level:two}. }
		\label{fig:level2:minor}
	\end{figure}
	
	(iv) This follows since the network $N_4$ on $\{x_1,x_2\}$ in Fig~\ref{fig:examples} 
	is level-4 but not planar as  $K_{3,3}$ is a minor of $U(N_4)$. Now  
	construct a phylogenetic network $N'_4$ on $X$ by replacing
	$x_1$ with any binary rooted phylogenetic tree on $\{x_1,x_3,..,x_n\}$.
	Then it is straight-forward to see 
	that $N'_4 \in \mathcal L_4(X)\setminus \mathcal P_{p}(X)$.
\end{IEEEproof}

In the last theorem we showed that all level-1 networks are outer planar.
In the second main result of this section, we show that for a level-2 network 
outer planarity of the network is equivalent to 
it being terminal planar, and characterize 
those level-2 networks that have this property.

\begin{thm}
	\label{thm:level:two}
	Suppose that $N$ is a level-2 phylogenetic network on $X$. Then the following statements are equivalent:
	\begin{itemize}
		\item[(i)] $N$ is terminal planar.
		\item[(ii)] $N$ is outer planar. 
		\item[(iii)] $N$ does not contain any subdivision of any of the four digraphs in Fig.~\ref{fig:level2:minor}.
		\item[(iv)] $U(N)$ does not contain a subdivision of $K_{2,3}$.		 
	\end{itemize}
\end{thm}

\begin{IEEEproof}
	Since  $(ii) \Rightarrow (i)$ clearly holds, it suffices to show that 
	$(i)\Rightarrow (iii)\Rightarrow (iv) \Rightarrow (ii)$. 
	
	$(i) \Rightarrow (iii)$: First assume that $N$ contains a subdivision of $H'_1$  in Fig.~\ref{fig:level2:minor}. Since $N$ is a binary network, it follows that there exists two elements $x_1$ and $x_2$ in $X$ such that for $i=1,2$, there exists a directed path from $v_i$ to $x_i$ which does not contain any directed edge in $H'_1$. 	
	Now an argument similar to the proof of Theorem~\ref{thm:level}(ii) (see also Fig.~\ref{fig:case1}) shows that $N$ is not terminal planar, a contradiction. The other three cases can be established in a similar manner.
	
	$(iii) \Rightarrow (iv)$: Assume that $U(N)$ contains a subdivision of $K_{2,3}$.
	We need to show that $N$ contains a subdivision 
	of one of the four  digraphs in Fig.~\ref{fig:level2:minor}.  
	Let $H$ be a biconnected component in $N$ such that $U(H)$ contains a 
	subdivision of $K_{2,3}$, which must clearly exist. 
	Then $H$ contains a subdivision of one 
	of the four digraphs $H_i$ in Fig.~\ref{fig:generator}. 
	We assume that $H$ contains a 
	subdivision of $H_2$; the other cases can be established in a similar manner.  
	
	We shall show that $H$ contains a subdivision of $H'_2$.
	To this end, label the vertices and edges in $H_2$ as in Fig.~\ref{fig:level2:cases}(A). 
	Since $U(H)$ contains a subdivision of $K_{2,3}$, there exists two vertices $u,v$ in $U(H)$ 
	such that there are three internally disjoint paths between them and each of these 
	three paths contains at least two edges.  Noting that $H'_2$ is obtained from $H_2$ 
	by subdividing the edge $e_2=\{u_1,u_2\}$ with the path $u_1,v_1,u_2$, it is 
	now sufficient to establish the following claim:
	
	{\bf Claim:} Under the above assumptions, $\{u,v\}=\{u_1,u_2\}$.	
	
	To establish the claim  we consider the following subcases 
	based on the size of intersection $\{u,v\}\cap V(H_2)$.
	
	Case (a): Neither $u$ nor $v$ is contained in $V(H_2)$. 
	Then both $u$ and $v$ are contained in edges of $H_2$ 
	and are added in the subdivision step when obtaining $H$ from $H_2$. 
	We have two subcases to consider. The first one assumes that both 
	$u$ and $v$ are contained in the same edge. We 
	assume that both of them are contained in $e_1$; the other cases are similar 
	(see  Fig.~\ref{fig:level2:cases}(B) for an illustration). 
	Then at most one path between $u$ and $v$ in $U(H)$ 
	does not contain the vertex $u_2$, and hence $U(H)$ 
	does not contain a subdivision of $K_{2,3}$, a contradiction.
	The second subcase assumes that $u$ and $v$ are contained in two 
	different edges. We assume $v$ is contained in $e_1$ and $u$ is contained in $e_4$; the
	other cases are similar (see  Fig.~\ref{fig:level2:cases}(C) for an illustration). 
	Then each path between $u$ and $v$ in $U(H)$ contains either $u_1$ or $u_2$, 
	and hence $U(H)$ does not contain a subdivision of $K_{2,3}$, a contradiction.
	
	Case (b): Precisely one of the vertices between  $u$ and $v$ is contained 
	in $V(H_2)$. Without loss of generality, we assume 
	$u\in V(H_2)$ and $v\not \in V(H_2)$. 
	Then there are 30 combinations to consider:  five choices
	for $u$ and six choices of $v$. Here we consider 
	the combination $u=u_3$ and $v$ is contained in $e_1$ (see  Fig.~\ref{fig:level2:cases}(D) for an illustration);
	the other combinations can be proved similarly. 
	For this combination, each path 
	between $u$ and $v$ in $U(H)$ contains either $u_1$ or $u_2$, 
	and hence $U(H)$ does not contain a subdivision of $K_{2,3}$, a contradiction.
	
	Case (c): Both vertices are contained in $V(H_2)$, that is, $\{u,v\}\subseteq V(H_2)$. 
	Since $\{u,v\}\not =\{u_1,u_2\}$, then an argument similar to the proof of Case (a) 
	shows that $U(H)$ does not contain a subdivision of $K_{2,3}$, a contradiction. 
	This completes the proof of the claim, and hence also the implication $(iii) \Rightarrow (iv)$. 		
	
	\begin{figure*}[ht]
		\center
		\scalebox{0.6}{\includegraphics{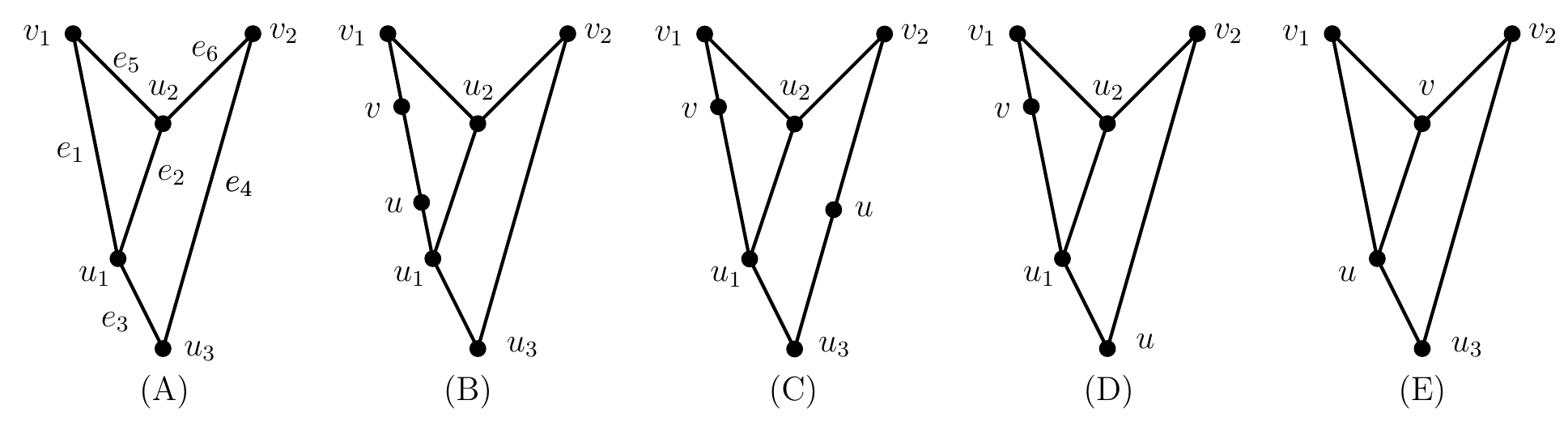}}
		\caption{ Digraphs for level-2 networks in the proof of~Theorem~\ref{thm:level:two}. }
		\label{fig:level2:cases}
	\end{figure*}
	
	$(iv) \Rightarrow (ii)$: Since $N$ is outer planar if and only if 
	$U(N)$ does not contain a subdivision of $K_{2,3}$ or $K_{4}$, 
	it remains to show that $N$ does not contain a subdivision of  $K_{4}$. 
	If this were not the case, then there must exist a biconnected component $B$ 
	of $U(N)$ such that $B$ contains a subdivision of  $K_{4}$. This 
	implies that we need to delete at least $3$ edges from $B$ to convert 
	it into a tree, in contradiction to the assumption that $N$ is level-2. 
\end{IEEEproof}

\noindent
{\bf Remark:} Note that the condition that $N$ is level-2 
is optimal in Theorem~\ref{thm:level:two} in the following sense:  The 
network $N_1$ in  Fig.~\ref{fig:examples} is level-3 and terminal planar, but not outer planar. 

\section{Regular networks}
\label{sec:regular}

In this section, we shall consider planarity of regular 
phylogenetic networks \cite{BS05} 
that are defined as follows. First, note that given any cluster system $\C$ 
on $X$, we can regard the Hasse diagram $H(\C)$ as a phylogenetic network on $X$, where 
the root is the vertex corresponding to the cluster $X$, and each 
element $x \in X$  corresponds to the singleton set $\{x\}$.
Conversely, given a phylogenetic network $\N=(V,A)$ on $X$, consider the
partial ordering $\preceq_\N$ on $V$ (or $\preceq$ for short)  
given by taking $u\preceq_\N v$ for $u,v\in V$ if there exists a 
directed path from $u$ to $v$ (which includes the special case that $u=v$).
The {\em cluster induced} by a vertex $v$ in $\N$, denoted by $\C(v)$, 
is the set of elements $x \in X$ such that $v\preceq x$ (so
$u\preceq_\N v$, $u,v \in V$ if and only if $\C(v) \subseteq \C(u)$), and 
the cluster system  $\C(\N)$ {\em induced by $\N$} consists of all 
clusters induced by the vertices in $\N$. 
A phylogenetic network $\N$ on $X$ is {\em regular} if $\N$ is isomorphic to $H(\C(\N))$~\cite[p.252]{S}. 

Interestingly, there is a special type of planar, regular 
phylogenetic network on $X$ that first appeared  sometime ago in the literature \cite{Diday}
(using different terminology).
Following \cite{K13}, a cluster system $\C$ of $X$ is a {\em prepyramid} 
if it can be represented as a family of intervals, that is, there exists a bijective 
mapping $f$ from $X$ to $[n]:=\{1,\dots,n\}$ such that $f(A)$
is an interval of $\{1,\dots,n\}$ for each cluster $A$ in $\C$, and a {\em pyramid} 
if it is both pre-pyramid and closed under intersection~\cite{Diday}.
Here we use the convention that singleton sets are also intervals.
We define a  {\em (pre)pyramid network} to be a phylogenetic network on $X$ of 
the form $H(\C)$ for $\C$ a (pre)pyramid on $X$.
In \cite{Diday} pyramid networks were 
implicitly defined and shown to be planar.
An example of such a network also appeared recently in \cite[Fig 2.]{BSH22}. 
\rev{Note that pyramids are closely related to the so-called circular split systems (see e.g.~\cite{K13}), the type of split systems that underpin the NeighborNet algorithm mentioned in the introduction. 
}

In this section, we shall show that terminal planar 
regular networks and pyramidal networks are in 
fact one and the same thing. To prove this, we shall use the following three 
results rephrased in our terminology: (F1) the poset induced by a planar digraph 
with one source and one sink is a lattice \cite[Theorem 5]{G95}, (F2) a lattice 
is upward planar if and only if it has dimension at most 2 \cite[Proposition 5.2]{K75}, and 
(F3) if $\C$ is a prepyramid on $X$, then 
the poset $(\C,\subseteq)$ has dimension at most 2 \cite[Theorem 3.6.1]{D41}. 

\begin{thm}\label{thm:regular}
	Suppose that $N$ is a regular phylogenetic network on $X$. Then $N$ is terminal planar if and only if $N$ is pyramidal. 
\end{thm}

\begin{IEEEproof}
	Let $N$ be a regular network on $X$. Denote its cluster system by $\C=\C(N)$ so that, since 
	$N$ is regular, $N\cong H(P(\C))$. Clearly the isomorphism between $N$ and $H(P(\C))$ 
	can be extended to give a digraph isomorphism between the completion $N^*$ of $N$ and 
	the Hasse diagram $H(P^*(\C))$, by mapping the sink vertex $t$ in $N^*$ to the vertex in $H(P^*(\C))$ 
	associated with the bottom element.
	
	For the `if' direction, suppose that $N$ is pyramidal, so that $\C$ is a pyramid.
	Then by Lemmas~\ref{lem:closed:lattice} and the fact (F3) above, 
	$P^*(\C)$ is a lattice 
	with dimension at most 2. It follows that $H(P^*(\C))$ is upward planar by the fact (F2) above. 
	Hence, as $N^*$ is isomorphic to  $H(P^*(\C))$, it follows that $N^*$  
	is upward planar, and so $N$ is terminal planar by Theorem~\ref{thm:upward}. 
	
	Now, for the `only if' direction, suppose that $N$ is terminal planar.
	By Theorem~\ref{thm:upward} $N^*$ is upward planar.
	As $N^*$ is isomorphic to  $H(P^*(\C))$, by 
	the fact (F1) above
	it follows that $P^*(\C)$ is a lattice.
	Therefore, $\C$ is closed under intersection, that is, 
	for any two vertices $u$ and $v$ in $N$ with $\C(u)\cap \C(v)\not =\emptyset$, 
	there exists a vertex $w$ in $N$ such that $\C(u)\cap \C(v)=\C(w)$. 
	
	Fix an upward planar drawing $\Gamma^*$ of $N^*$, and 
	let $\Gamma$ be the terminal planar drawing of $N$ 
	induced by $\Gamma^*$, that is, $\Gamma$ is obtained from $\Gamma^*$ 
	by removing the node corresponding to the sink vertex in $N^*$ and all arcs incident with it. 
	Consider the ordering of
	the elements in $X$ obtained from the drawing $\Gamma$ of $N$ by following the outer-face of $\Gamma$ 
	in a clockwise direction starting from the node associated with the root of $N$. 
	This gives a bijective map $f$ from $X$ to $\{1,\dots,n\}$, where $n=|X|$. 
	For simplicity, we may further assume that the elements in $X$ are enumerated 
	in a way such that $f(x_i)=i$ holds for all $1\le i \le n$.
	We claim that for each cluster $A$ in $\C$, $f(A)$ is an interval of $\{1,\dots,n\}$ which, as
	$\C$ is closed under intersection, will complete the proof of the theorem. 
	
	To establish the claim, consider an arbitrary vertex $v$ in $N$. Without loss of generality, we suppose that $a<b$ are two numbers in $\{1,\dots,n\}$ 
	with $a<b-1$ and  both $x_a\preceq v$ and $x_b\preceq v$ 
	hold for $x_a:=f^{-1}(a)$ and $x_b:=f^{-1}(b)$, as otherwise the claim clearly holds. It suffices to show 
	that for each $c$ with $a<c<b$ we have $x_c\preceq v$ for 
	the leaf $x_c:=f^{-1}(c)$. This clearly holds if $v$ is the root 
	of $N$, and hence we may assume that $v$ is not the root for the remainder
	of the proof of the claim.
	
	Fix two directed paths $H_{a}$ and $H_b$ in $N^*$ between $v$ and $x_a$ and $x_b$, respectively.
	Furthermore, let $e_a=(x_a,t)$ and $e_b=(x_b,t)$ denote the two directed
	edges connecting $x_a$ and $x_b$ to the sink element $t$, respectively. 
	Without loss of generality, we may further assume that $v$ is the only 
	common vertex contained in both $H_{a}$ and $H_b$, as otherwise we 
	can replace $v$ by the lowest vertex in $H_a$ that is also contained in $H_b$.  
	
	Let $L_a=\Gamma^*(H_a)$ and $L_b=\Gamma^*(H_b)$ be the 
	curves in the drawing $\Gamma^*$ corresponding to $H_a$ and $H_b$, respectively. 
	Then the  curve $L=L_a\cup L_b\cup \Gamma^*(e_a) \cup \Gamma^*(e_b)$ forms a 
	closed Jordan curve in the plane.  
	By the Jordan Curve Theorem (see e.g.~\cite[Theorem 4.1.1]{Diestel}), 
	it follows that $L$ partitions the plane into  
	precisely two connected components, one containing the node 
	corresponding to root $\rho$, and the other containing the node corresponding to vertex $x_c$.
	
	Now, fix a path $H_c$ in $N^*$ from the root $\rho$ to $x_c$. 
	Then, as  $L$ partitions the plane into  
	precisely two connected components, $\Gamma^*(H_c)$ must intersect at least one
	node in $L$. That is, there exists a vertex $u$ in $\N^*$ 
	such that $\Gamma^*(u)\in L\cap \Gamma^*(H_c)$. Since $\Gamma^*$ 
	is an upward planar drawing of $\N^*$, we can 
	further assume $u\not \in \{x_a,x_b,t\}$.  
	Since  $v   \preceq u \preceq x_c$, it follows that $x_c\in \C(v)$, which 
	completes the proof of the claim.
\end{IEEEproof}

Note that there are regular networks on $X$ 
can have $2^{|X|}-1$ vertices (see, e.g.~\cite[Section 10.3.4]{S}), 
but by Theorem~\ref{thm:regular}, the 
number of vertices in any terminal planar regular network is less than or 
equal to ${|X| \choose 2}$, 
the number of intervals of $\{1,2,\dots,|X|\}$.
Thus by Theorem~\ref{thm:algorithm} we have:

\begin{cor} 
	If $N$ is a regular phylogenetic network on $X$, then we can check in $O(|X|^2)$ if $N$ is terminal planar.
\end{cor}

\begin{figure}
	\center
	\scalebox{0.5}{\includegraphics{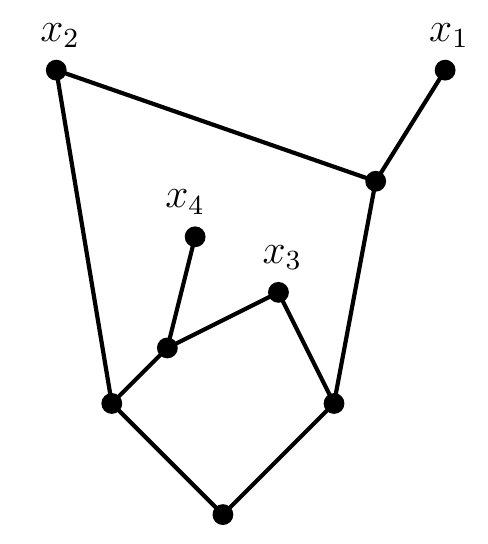}}
	\caption{  A regular network that is a prepyramid network and upward planar, but not terminal planar.}
	\label{fig:prepyramid}
\end{figure}

\noindent
{\bf Remark:} Consider the phylogenetic network $N$ on $X=\{x_1,x_2,x_3,x_4\}$ in Fig.~\ref{fig:prepyramid}. 
Then since 
\setlength{\arraycolsep}{0.0em}
\begin{eqnarray*}
\C(N) &=& \{ \{x_1,x_2,x_3,x_4\}, \{x_1,x_2,x_3\}, \{x_2,x_3,x_4\}, \\
&&\,\{x_1,x_2\}, \{x_3,x_4\}, \{x_1\}, \{x_2\}, \{x_3\},\{x_4\}\},
\end{eqnarray*}
\noindent
it is straight-forward to see that $N$ is a regular and that $\C(N)$
is a prepyramid. Moreover, $N$ is {\em not} a pyramid network because the 
intersection $\{x_1,x_2,x_3\} \cap \{x_2,x_3,x_4\}=\{x_2,x_3\}$ 
is not contained $\C(N)$. Thus, this gives an example
of a prepyramid $\C$  such that the phylogenetic network 
$H(\C)$ on $X$ is upward planar (in view of Fig.~\ref{fig:prepyramid}), 
but not terminal planar (interestingly, in this example the poset $P(\C(N))$ has dimension 2).

\section{Discussion}
\label{sec:discussion}

In this paper, we have considered the concept of 
planar phylogenetic networks, and 
shown that the theory and algorithms coming from the area of planar digraphs
can be very useful for handling and understanding these structures.

There remain several open questions. For example:
\begin{itemize}
	\item \rev{Further elucidate planarity for additional classes of rooted phylogenetic networks, such as galled and tree-based networks (see e.g.~\cite{zhang2019clusters} for a recent review).}	
	\item Given a planar phylogenetic network on $X$, 
	let $\widetilde{N}$ be the digraph obtained from $N$ by adding 
	an additional arc $(x,\rho)$ for each element $x$ in $X$ if $(\rho,x)$ is not already an arc in $N$.
	If $N$ is terminal planar, then it is straight-forward to see that
	$\widetilde{N}$ is planar, but does the converse hold?
	
	\item Given a normal network $N$, we define $Comp(N)$ 
	to be the network which is obtained from $N$ by collapsing 
	any arc $(u,v)$ in $N$,  where $u$ has outdegree 1 and $v$ has indegree 1.
	Note that that $Comp(N)$ must be regular (see e.g. \cite[p.253]{S}).
	Is $N$ is terminal planar if and only $Comp(N)$ is terminal planar?
	
	\item Is there a prepyramid  cluster system $\mathcal C$ such 
	that $H(\mathcal C)$ is not upward planar?
	
	\item Although there are some general algorithms for drawing 
	planar, upper planar, terminal planar and outer planar networks (Theorem~\ref{thm:algorithm}), are there more 
	specific algorithms for drawing special types of planar phylogenetic networks
	such as tree-child networks?
\end{itemize}

More generally, the phylogenetic networks that we have 
considered in this paper did not have arc weights. It would
be interesting to understand how arc weights might affect 
our results, especially when we want to make 
a straight-line drawing where the length of an
arc is proportional to its weights. A special case that could be 
considered first are  
temporal phylogenetic networks which incorporate a 
natural vertical time axis which provides timings for
past evolutionary events (see e.g. \cite[p. 250]{S}). This
concept appears to be related to upward planarity, and it 
would be interesting to further investigate this relationship. Moreover, in 
general as the theory of phylogenetic networks continues to  
grow it could be worthwhile to develop  new algorithms 
to produce planar phylogenetic networks from biological data.


%



\ifCLASSOPTIONcaptionsoff
  \newpage
\fi



%

\bibliography{pnets}{}
\bibliographystyle{IEEEtran}

\end{document}